\documentclass[twocolumn,amsmath,amssymb,pre]{revtex4}
\usepackage{graphicx}% Include figure files
\usepackage{dcolumn}% Align table columns on decimal point
\usepackage{bm}% bold math
\newcommand{\pp}{\mathit{P}}
\newcommand{\bp}{\bar{P}}
\newcommand{\by}{\bar{y}}

\newcommand{\expc}[1]{\left<#1\right>}
\newcommand{\dlangle}{\left\langle\!\left\langle}
\newcommand{\drangle}{\right\rangle\!\right\rangle}
\newcommand{\cum}[1]{\dlangle{#1}\drangle}

\begin{document}

\title{Statistical physics of power fluctuations in mode locked lasers}
\author{Oded Basis, Omri Gat}

\affiliation{Racah Institute of Physics, Hebrew University, Jerusalem 91904, Israel }

\date{\today}
%\begin{document}
%\maketitle
\begin{abstract}
We present an analysis of the power fluctuations in the statistical steady state of a passively mode locked laser. We use statistical light-mode theory to map this problem to that of fluctuations in a reference equilibrium statistical physics problem, and in this way study the fluctuations non-perturbatively. The power fluctuations, being non-critical, are Gaussian and proportional in amplitude to the inverse square root of the number of degrees of freedom. We calculate explicit analytic expressions for the covariance matrix of the overall, pulse and cw power variables, providing complete information on the single-time power distribution in the laser, and derive a set of fluctuation-dissipation relations between them and the susceptibilities of the steady-state quantities.
\end{abstract}
\maketitle

\section{Introduction}
Passive mode locking is a ubiquitous tool for creating ultra-short pulses, whose duration can be as short as a few femtoseconds. It is achieved by the simple device of placing a saturable absorber in a multi-mode laser cavity, that effectively amplifies optical waveform regions of high instantaneous power, thereby rendering cw waveforms modulationally unstable \cite{haus,kutz}. However, in contrast with dynamical model of passive mode locking, in experimental systems with a fast saturable absorber there exists a threshold for self-starting of mode locking \cite{haus-apm2,Spielmann}. The decohering noise sources present in the laser cavity are an inevitable impediment to the dynamical process of pulse buildup \cite{HausIppenReflections, MenyukSelfStarting, Krausz, Krausz2}. Therefore, the self-starting problem could only be resolved when a  \emph{statistical} theory of interacting laser light modes subject to noise  was put forward in \cite{GFPRL}. In statistical light-mode dynamics (SLD) the optical waveform is treated as a random function, and the mode locking phenomenology is recovered as the ``thermodynamics'' of the SLD system. It can then be shown \cite{GFPRL,GGF04,parabolic} that the noise injects \emph{entropy} into the cavity, and that the entropy of the cw state is higher than that of the mode locked state. The onset of passive mode locking therefore obtains the significance of a first order phase transition between the disordered cw phase and the ordered pulse phase. SLD has been further applied to predict and study several additional effects in passively mode locked lasers, including critical phenomena in light \cite{critical-PRL05}, multi-pulse phase transitions \cite{steps,steps07}, noise-activated mode locking \cite{escape}, and mode locking in random lasers \cite{randomlasers}.

At the same time, noisy models of passive mode locking were often considered, not for the purpose of understanding the mode locking transition, but to study the temporal fluctuations of the cavity waveform \cite{haus-mecozzi}. The presence of cavity noise perturbs the output of a laser from a perfectly periodic train of pulses, degrading the pulse-to-pulse coherence, temporal periodicity, and the pulse shape. In applications these fluctuations are usually obstructive, and one of the main goals of the study of fluctuations has been to understand how the fluctuations can be reduced. Recently the importance of understanding and controlling fluctuations has significantly increased, because the pulse fluctuations are the limiting factor in the accuracy of the ultra-precise clocks based on frequency combs \cite{nature,cundiff,fibercomb}. 

The standard theoretical approach to pulse fluctuations is based on perturbation theory, an apparently reasonable assumption, since the noise power entering the cavity is much smaller than the coherent power \cite{haus-mecozzi,paschotta,mlld,menyuk1,menyuk2}. However, this argument is valid only for a single pass (or few passes) through the amplifier. In a laser that is based on a dynamical equilibrium and feedback, the noise power accumulates \cite{haus}, and can no longer be considered small. In fact a theory of fluctuations based entirely on perturbation theory can only capture transients. This difficulty has been recently recognized by \cite{mlld,menyuk1} who demonstrated that the analysis of fluctuations must also take into consideration the gain dynamics. Therefore, a consistent theory of steady-state pulse fluctuations must include cw power variables in a statistical framework like SLD.

The analysis of pulse fluctuations in SLD is quite similar to that of statistical fluctuations in equilibrium statistical mechanics \cite{landau}. When there is no broken symmetry, thermodynamic quantities fluctuate around the equilibrium values. The distribution of the fluctuations is Gaussian, and their strength tends to zero in the thermodynamic limit being inversely proportional to the square root of the number of degrees of freedom. Thermodynamic quantities associated with broken symmetries undergo diffusion whose rate is also inversely proportional to the number of degrees of freedom. There exist proportionality relationships (fluctuation-dissipation relations) between correlation functions of the fluctuating quantities and thermodynamic susceptibility and transport coefficients. The thermodynamic quantities associated with a single pulse in a passively mode locked laser are the pulse power, frequency, timing and phase, of which the last two are symmetry breaking quantities. 

In this paper we focus on the \emph{power} fluctuations, since it is the pulse power that is directly coupled to the noise power and entropy through the amplifier gain \cite{mikatz,haus,dogru}, and the power fluctuations exhibit therefore the richest behavior. 
The power fluctuations are studied in the framework of a simplified model of the mode locking dynamics, where the cavity is divided into $N$ intervals with the width of a pulse, $N$ being the number of active modes in the cavity, and each interval is represented by a single complex degree of freedom. The model has been introduced and derived from the Haus master equation in \cite{GGF04}, and it has been demonstrated  to yield results that closely approximate those obtained from the master equation on one hand \cite{parabolic}, and have large experimentally predictive content on the other hand \cite{steps}. The model, termed here the coarse-grained model of passive mode locking, is defined in section \ref{sec:cg}, its thermodynamic properties are reviewed, and the thermodynamic power observables are defined and listed. The expectation values of the thermodynamic observables define the steady state of the laser, characterized by the overall power, the pulse power, and the cw power. The values of the thermodynamic quantities are calculable from an exact mean field theory, and depend on the single dimensionless parameter $\gamma=\frac{\alpha\bp^2}{T}$, where $\alpha$ is the absorber nonlinearity, $\bp$ is the overall cavity power, and $T$ is the noise power injection rate. In particular the pulse power is $\by\bp$, where $\by$ is a function of $\gamma$ only. The role of $\gamma$ in SLD is analogous to inverse temperature in statistical mechanics, in that the ordering transition occurs, and the pulse power increases, as $\gamma$ increases \cite{GGF04}.

Since the fluctuations are Gaussian, they are completely defined in terms of the covariance matrix of the thermodynamic observables, which is calculated explicitly in Sec. \ref{sec:fluctuations}, the main section of the paper. It is shown that,  in addition to $\gamma$, the fluctuations depend on a second dimensionless parameter $\eta=\bp|g'(\bp)|$, where $g(P)$ is the saturated overall net gain. $\eta$ can be thought of as the gain elasticity, as it determines the restoring force of the gain saturation mechanism in establishing the steady state of the laser. If $\eta$ becomes very large, the fluctuations in the overall power tend to zero, but there still remain pulse power fluctuations, as power is randomly passed back and forth between the pulse and the cw degrees of freedom. The covariance matrix is first derived directly from the invariant measure. There are overall seven independent elements in the covariance matrix, and the results are summarized in table \ref{tab:cum} (for entries depending at most on a single cw power observable), and Eqs. (\ref{eq:cXX}--\ref{eq:cxx'}), and convey full information on the single time statistical properties of the fluctuations. Next, a set of fluctuation-dissipation relations is derived, which are used to express several of the covariance matrix elements as derivatives of steady state thermodynamic quantities, thereby providing an independent verification of their values. Finally, the covariance matrix is also calculated in the much simpler case, where the laser is in the disordered phase, i.e. when it is operating in cw; there it is shown that the fluctuations do not depend on $\gamma$, so that they are unaffected by the saturable absorber. The last section of the paper presents our conclusions and outlook.

\section{The coarse-grained model of passive mode locking}\label{sec:cg}
The effective number of degrees of freedom in a multimode laser cavity is determined by the bandwidth of the gain medium, which determines the pulse width $\tau_p$. Thus, a cavity of round-trip time $\tau_R$ has $N=\tau_R/\tau_p$ independent degrees of freedom, which can be thought of as either $N$ Fourier modes, or, in real space, as $N$ independent complex field amplitudes $\psi_n$, each one representing the average value of the optical field an in interval of length $\tau_p$.

In this picture statistical light mode dynamics is described by $N$ coupled ordinary differential equations \cite{GFPRL,GGF04}
\begin{equation}\label{eq:cg}
\partial_t\psi_n=\alpha|\psi_n|^2\psi+g\psi_n+\eta_n\ ,
\end{equation}
where $\alpha$ is the coefficient of saturable absorption, $g$ is the overall net gain, and the $\eta_n$'s are independent (complex) centered Gaussian white noise processes with covariance functions $\left<\eta_n^*(t)\eta_m(t')\right>=2T\delta(t-t')$. 

The equations (\ref{eq:cg}) have to be supplemented by an equation that expresses the gain coefficient in terms of the field variables using the physical gain saturation mechanism. In the case of slow gain saturation considered here, the gain coefficient can be assumed to be a function of the \emph{overall power} $\pp=\frac1N\sum_n|\psi_n|^2$. In many studies the gain saturation function is modelled by $g(\pp)=\frac{g_u}{1+\pp/P_s}$ (for a constant $P_s$) \cite{haus,kutz}. However, it is known \cite{kutz,kutz-stability} that the mode locking dynamics with a $|\psi|^2\psi$ nonlinearity is globally unstable with this kind of gain saturation function. Since, furthermore, the steady state properties of the optical waveform depend only on the local behavior of $g(\pp)$ near the operational power $\bp$ (defined precisely below), we will keep an arbitrary gain saturation function $g(P)$, that is only assumed to have the necessary properties for a stable mode locking state to exist; in this way we gain in generality without complicating the analysis.

The equations of motion (\ref{eq:cg}) can be written in a variational form \cite{GGF04}
\begin{equation}
\partial_t\psi_n=-\frac{\partial H}{\partial\psi_n^*}+\eta_n\ ,
\end{equation}
where $H$ is the Lyapunov functional defined by
\begin{equation}
H[\psi]=-\frac\alpha{2N}\sum|\psi_n|^4+NTu(\pp)\ ,
\end{equation}
and $u(\pp)$ is the dimensionless gain potential, defined by $-NTu'(\pp)=g(\pp)$. As shown in \cite{GGF04}, the $N$ scaling of the gain potential is necessary to obtain a good thermodynamic limit as $N\to\infty$.

In this paper we study the steady state statistical properties of the SLD coarse-grained model. As a white-noise forced variational system, the system (\ref{eq:cg}) reaches a statistical steady state with a Gibbs-like distribution
\begin{equation}\label{eq:rho}
\rho[\psi]=\frac{e^{-H[\psi]/T}}{Z}\ ,
\end{equation}
where $Z$ is the partition function
\begin{equation}
Z=\int[d\psi]e^{-H[\psi]/T}\ .
\end{equation}
Steady-state expectation values of field functionals $A[\psi]$ are given as usual by
\begin{equation}\label{eq:<A>}
\left<A\right>=\int[d\psi]\rho[\psi]A[\psi]\ .
\end{equation}
This is the starting point for the calculation of the mean values of macroscopic observables and their (co)variance.

In previous studies \cite{GFPRL,GGF04,steps,critical-PRL05} the thermodynamics of this system, i.e.\ the behavior of macroscopic observables when $N$ is large, has been studied in detail, and the main results are as follows: The overall power $\pp$ is self-averaging and reaches a value $\bp$ that is determined by a balance of the saturable absorption nonlinearity, noise entropy, and gain saturation. Once $\bp$ is determined, $u(P)$ plays no further part in the determination of the thermodynamics. Moreover, $\bp$ itself appears in the thermodynamics, except as an overall power scale, only in the dimensionless combination $\gamma=\frac{\alpha\bp^2}{T}$ that plays here a role similar to inverse temperature in equilibrium statistical physics. 
The thermodynamic phase diagram consists of two phases, a mode locked phase with a single pulse whose power $\bp\by$, $\frac12\le\by<1$, is also self-averaging in the thermodynamic limit, and a cw phase where the intracavity power is distributed evenly between all the degrees of freedom. 
 
The thermodynamics is solvable using an exact mean field theory. The mean-field free energy (Landau function) is
\begin{equation}\label{eq:fpy}
f(P,y)=\frac\gamma2 y^2+\log (P(1-y))-u(P)\ .
\end{equation} 
The value of $\by$ for a given overall power is determined by the equation $\partial_y f(P,\by)=0$ whose solution is
\begin{equation}\label{eq:bary}
\bar y=\frac12+\sqrt{\frac14-\frac1\gamma}\ (\gamma>4),
\end{equation}
that does not depend directly on $u$ and depends on $P$ only through the combination $\gamma$ as explained above. Therefore, if only the overall power $P$ is fixed and the pulse power is let to assume its equilibrium value, the free energy becomes
\begin{equation}
\phi(P)=f(P,\by(P))\ .
\end{equation}
The steady state overall power is now determined by the equation $\partial_P\phi=0$ that with Eq.\ (\ref{eq:bary}) gives
\begin{equation}\label{eq:gp}
\gamma\bar y=\bp u'(\bp)\ .
\end{equation}
This equation also determines also the overall net gain $g$, since $u'$ is proportional to it. The gain balance requirement is an alternative method for reaching (\ref{eq:gp}), based on dynamics \cite{mikatz}.

The values of $\bp$ and $\by$ derived above describe a mode locked state that exists for every $\gamma>4$, whose free energy is $F=f(\bp,\by)$. In addition, for any $\gamma$ there exists a cw state that can be viewed formally as a state with zero-power pulse, so that the mean-field free energy in the cw phase is $\phi_\text{cw}(P)=f(P,0)$, and, the overall power in the cw state is determined by the condition $\partial_P\phi_\text{cw}(\bp_\text{cw})=0$, that is,
\begin{equation}\label{eq:gpcw}
\bp_\text{cw}u'(\bp_\text{cw})=1
\end{equation}
Of the two states, the statistically stable is the one with the lower free energy, and the other is metastable, although in general its lifetime is very large \cite{escape}. The two states exchange stability when $\gamma=\gamma^*$, above which the mode locked phase becomes stable. $\gamma^*$ is always greater than 4, but its precise value depends on the form of the gain saturation function $u$. For very strong gain saturation, i.e.\ when $\eta$ tends to infinity, $\gamma^*$ approaches a value close to $4.91$ \cite{GGF04}.

The overall, pulse and cw power variables are strictly equal to the thermodynamic quantities shown above only in the limit $N=\infty$. When $N$ is large but finite, their actual values deviate slightly from the thermodynamics, and fluctuate in time. These fluctuations are observed as a small randomness in the macroscopic observables. Since the mode system is not critical, the fluctuation statistics are well-described by a (multi-variate) Gaussian distribution \cite{landau}. The nature of the coarse-grained model and the equilibrium-like approach are well suited to study the single-time power fluctuations. Since a macroscopic share of the overall power resides in the cw background even in the mode locked phase, the macroscopic observables include, in addition to $\pp$ and the pulse power $Y$, partial continuum power variables $X_a=\frac1N\sum\limits_{\text{interval}}|\psi_n|^2$, where the sum is taken on an interval containing $a N$ degrees of freedom, $0<a\le1$. In the interest of examining continuum cross-correlations we will also define complementary continuum power variables $X'_b$ defined similarly to the $X_a$ variables on  mutually exclusive intervals of length $b N$.

The expectation values of all these macroscopic variables reach  finite positive limit 
\begin{equation}
\expc{P}=\bp\ ,\quad\expc{Y}=\bp\by\ ,\quad\expc{X_a}=a\bp(1-\by)\ ,
\end{equation}
 as $N\to\infty$,
while their (co)variance are $O(1/N)$, so that the relative size of the fluctuations is $N^{-1/2}$ as expected.  Our main result is the derivation of covariance matrix of these observables, to which we now turn.

\section{Fluctuation statistics of the power observables}\label{sec:fluctuations}
\subsection{The covariance matrix in the mode locked phase}\label{sec:cov}
In this subsection it is assumed that the system is in the mode locked phase, that is, there exists a single degree of freedom whose power is close to $N\bp\by$, where $\bp$ and $\by$ are the overall power and relative pulse power in the thermodynamic limit, as defined above. This assumption implies that $\gamma=\frac{\alpha\bp^2}{T}>4$. When $\gamma<\gamma^*$ this state is metastable, but except for parameters very close to the boundary of the pulse stability region, the metastable state is very long-lived, and the fluctuation statistics are well-defined. 

\begin{table*}[ht]
\begin{tabular}{c|ccc}
$N\bp^{-2}\cum{\cdot,\cdot}$& $P$ & $Y$&$X_a$\\[1mm] \hline
$P$        & $\displaystyle\frac{(2\by-1)}{(2\by-1)\eta-\gamma\by} $
           & $\displaystyle\frac{\by}{(2\by-1)\eta-\gamma\by} $
           & $\displaystyle-\frac{a}{(2\by-1)\by\eta-\gamma\by^2}$ \\[3mm]
$Y$    &   & \quad$\displaystyle\frac{1}{2\by-1}\Bigl(\frac{\by^2}{(2\by-1)\eta-\gamma\by} + \frac{1}{\gamma^2\by}\Bigr)$
           & \quad$\displaystyle-\frac{a}{2\by-1}\Bigl(\frac{1}{(2\by-1)\eta-\gamma\by}+\frac{1}{\gamma^2\by}\Bigr)$  \\
\end{tabular}
\caption{\label{tab:cum}The independent elements of the covariance matrix of the thermodynamic power observables, containing at most one cw power factor. The dimensionless entries in the table should be multiplied by $\frac{\bp^2}{N}$ to obtain the physical values. They are expressed in terms of the two dimensionless parameters $\gamma$, the inverse temperature, and $\eta$, the gain elasticity. As $\eta\to\infty$, the elements in the first row tend to zero, while those in the second row have a nonzero limit.}
\end{table*}

The elements of the covariance matrix are the second order cumulants of the macroscopic observables, i.e., combinations of expectation values (with respect to the invariant measure $\rho[\psi]$) of the type $\cum{AB}=\expc{AB}-\expc{A}\expc{B}$ where $A$ and $B$ are two, not necessarily distinct, of the thermodynamic observables $P$, $Y$, $X_a$ and $X'_b$ defined above. Using relations between these observables, and the phase symmetry of the cw degree of freedom (see below), all such cumulants can be expressed in terms of these containing at most a single factor of $X_a$. Expectation values linear in $X_a$ are equal to $Na$ times these expectation values with $X_a$ replaced by $x=|\psi_n|^2$, for $n$ an arbitrary cw degree of freedom, so that we need to evaluate expectation values of the form $\left<P^nY^mx^k\right>$, where $n,m,k$ are nonnegative integers whose sum is at most 2, and $k\le1$.
As shown in the appendix, such expectation values can be expressed
$\left<P^nY^mx^k\right>=I_{nmk}/I_{000}$, where
\begin{equation}\label{eq:inmk}
I_{nmk}=\int dP P^{n+m+k}\int dy y^m (1-y)^k e^{Nf(P,y)}\ ,
\end{equation}
where $f(P,y)$ is the free energy defined above. The large $N$ asymptotics of these integrals can be derived with the method of steepest decent (see appendix) and it has the general form
\begin{equation}\label{eq:inmksim}
I_{nmk}\sim e^{Nf(\bar P,\bar y)}\bar P^{n+m+k}\bar y^m(1-\bar y)^k\Bigl(1+\frac{\tilde I_{nmk}(\bar P,\bar y)}{N}\Bigr)\ ,
\end{equation}
where $\bar P$ and $\bar y$ and $f$ are the thermodynamic quantities defined above. Here $\sim$ signifies asymptotic equality up to an $O(1)$ prefactor that cancels in the calculations of expectation values.

In calculating the cumulants, the leading $O(1)$ terms in $I_{nmk}$ cancel, so that
\begin{multline}
\cum{P^nY^mx^k}=\\\qquad=\bar P^{n+m+k}\bar y^m(1-\bar y)^k(\text{linear combination of $\tilde I$'s})\ .
\end{multline}
The technical but straightforward calculation of the functions $\tilde I_{nmk}(\bar P,\bar y)$ is outlined in the appendix, and we proceed to state the results.

It is possible to divide the power fluctuations into two parts, fluctuations that occur with a \emph{given} overall power as a result of the random redistribution of this power between the various degrees of freedom, and those that are a consequence of the fluctuations in the overall power. These two contributions correspond to the $O(N^{-1})$ terms in the two integrations in (\ref{eq:inmk}). The first contribution that is generated by the $y$ integration is inversely proportional to $|\partial_y^2f|=\gamma^2(2\by-1)\by$, and is independent of the gain saturation function. It represents the dominant term in the fluctuations when gain saturation is so strong that overall power fluctuations are suppressed, that is, when $\eta\to\infty$. This term is absent in covariances that involve $P$. The second contribution, generated by the $P$ integration is inversely proportional to $|\partial_P^2\phi|^{-1}=\eta+\frac{\gamma\by}{1-2\by}$. Using this information we can calculate the covariance matrix elements involving at most one factor of $X$. The values of the five independent matrix elements are listed in table \ref{tab:cum}.

\begin{figure}
\includegraphics[width=8cm]{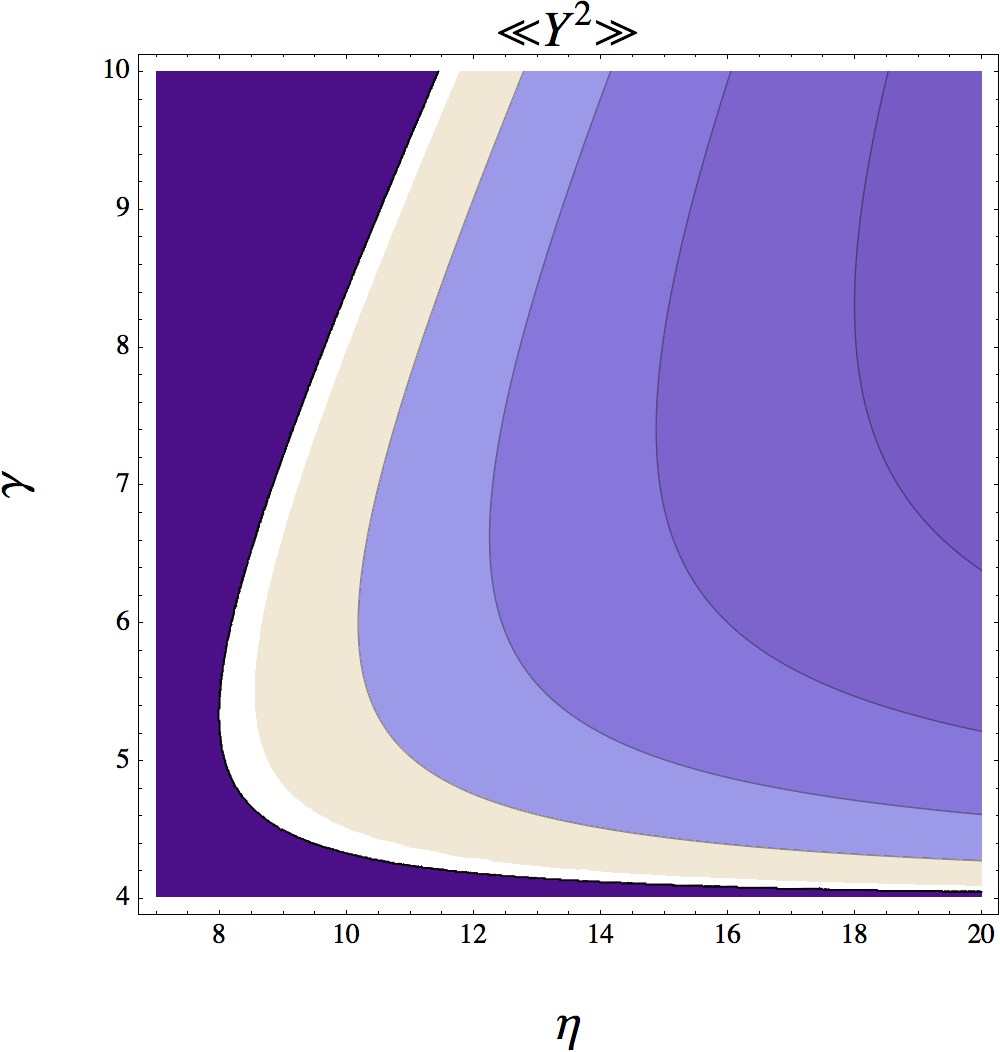}
\caption{\label{fig:cyy} A contour plot of the pulse power fluctuations variance $\cum{Y^2}$ as a function of the dimensionless parameters $\gamma$ (inverse temperature) and $\eta$ (gain elasticity). Lighter shades represent stronger fluctuations and darker weaker fluctuations, except the dark blue region in the left part of the figure, that designates parameter values where the pulse is unstable. Contour lines represent the values (from right to left) 0.2, 0.3, 0.5, and 1, times $\frac{\bp^2}{N}$.
}
\end{figure}

The qualitative dependence of the power fluctuations on the two dimensionless parameters is demonstrated in Fig.\ \ref{fig:cyy} that shows the normalized value of $\cum{Y^2}$ as a function of $\eta$ and $\gamma$. It is evident that, as expected, the typical fluctuations are proportional to $\frac{\bp}{\sqrt N}$, with an $O(1)$ coefficient, when the parameters are such that the pulse is stable. The power fluctuations grow when the parameters approach the boundary of the of the region of stability, where the amplitude of fluctuations diverges. For a fixed value of $\eta$, this can happen in two manners: When $\gamma$ is too small, the pulse becomes unstable with respect to noise buildup, and breaks down with the laser reverting to the cw state, and when $\gamma$ is too large, the gain saturation is too weak to prevent the saturable-absorber driven growth of a pulse perturbation. Thus, for $\eta>8$ the fluctuations first decrease as function of increasing $\gamma$, and then start increasing as the upper stability boundary is approached. For fixed $\gamma$ on the other hand, the fluctuations decrease monotonically for increasing $\eta$, reaching a finite value as $\eta\to\infty$.

We now consider the remaining elements of the covariance matrix elements, containing two cw factors. These are
\begin{align}
\cum{X_a^2}&=\frac{a}{N}\cum{x^2}+a^2\cum{xx'}\label{eq:cXX}\\
\cum{X_a X_b'}&=ab\cum{xx'}\label{eq:cXX'}
\end{align}
where $x=|\psi_n|^2$, $x'=|\psi_m|^2$, for arbitrary cw degrees of freedom $n\ne m$. Since the $x$ variables are microscopic, the variance $\cum{x^2}$ is $O(1)$ rather than $O(N^{-1})$ and contributes significantly in Eq.\ (\ref{eq:cXX}). Its statistics can be deduced from the fact that the $\psi_n$'s are (approximately) Gaussian random variables, and that $\expc{\psi_n}=0$ for cw degrees of freedom because of the phase symmetry of the equations of motion (\ref{eq:cg}), so that $\expc{x^2}=\expc{|\psi_n|^4}=2\expc{|\psi_n|^2}^2=2(\bp(1-\by))^2$, and therefore
\begin{equation}
\cum{x^2}=(\bp(1-\by))^2\ .
\end{equation}

The covariance $\cum{xx'}$ is actually linearly dependent on the covariances already calculated, as can be observed by calculating the cumulants of both sides the identity $\pp-Y=X_1$, giving
\begin{align}
\cum{xx'}&=\cum{P^2}-2\cum{PY}+\cum{Y^2}-\frac{1}{N}\cum{x^2}\nonumber\\
&=\frac{(1-\by)^2}{2\by-1}\Bigl(\frac{1}{(2\by-1)\eta-\gamma\by}+\frac{1-\by}{2\by-1}\Bigr)\frac{\bp^2}{N}\ .\label{eq:cxx'}
\end{align}

\subsection{Fluctuation-dissipation relations}
In equilibrium statistical physics, susceptibilities are linearly related to fluctuation covariances and transport coefficients are linearly related to diffusion coefficients \cite{callen} by the property of detailed balance. These relations can be easily derived directly from the Gibbs distribution. In our case, we write the Lyapunov functional as $H[\psi]=N(\alpha R[\psi]+\beta S[\psi])$, with $R[\psi]=-\frac{1}{2N^2}\sum_n|\psi_n|^4$ and $S[\psi]=T u(P)$, and let the thermodynamic quantities and fluctuation coefficients depend parametrically on $\alpha$ and $\beta$, with the physical quantities obtained for $\beta=1$. In the mode locked phase the partition function is given by $\log Z=NF+O(1)$, so that
\begin{align}
\partial_\alpha F&=-\frac{\expc{R}}{T}=\frac{(\bp\by)^2}{2T}\ .\label{eq:daf}\\
\partial_\beta F&=-\frac{\expc{S}}{T}=-u(\bp)\ .\label{eq:dbf}
\end{align}
The left-hand equality in Eqs. (\ref{eq:daf}--\ref{eq:dbf}) follows by differentiation of the partition function under the integral sign, and the definition (\ref{eq:<A>}) of the expectation values, and the right-hand equality from direct  differentiation of Eq.\ (\ref{eq:fpy}). The results obtained for $\expc{R}$ and $\expc{S}$ are actually special cases of the fact that $\expc{k(P,Y,X_a)}=k(\bp,\bp\by,a\bp(1-\by))$ for any function $k$ of the thermodynamic observables in the thermodynamic limit, that holds since they are self-averaging. Note that $R\sim\frac{Y^2}{2}$ in the thermodynamic limit, since the cw terms in the sum give an $O(\frac1N)$ contribution.

Taking another derivative of the free energy then gives the second order cumulants,
\begin{align}
\partial_\alpha^2 F_\text{ml}&=\frac{\cum{R^2}}{T^2}=-\frac{\partial_\alpha\expc{R}}{T}\label{eq:daaf}\\
\partial_\alpha\partial_\beta F_\text{ml}&=\frac{\cum{RS}}{T^2}=-\frac{\partial_\alpha\expc{S}}{T}=-\frac{\partial_\beta\expc{R}}{T}\label{eq:dabf}\\
\partial_\beta^2 F_\text{ml}&=\frac{\cum{S^2}}{T^2}=-\frac{\partial_\beta\expc{S}}{T}\label{eq:dbbf}
\end{align}
The right-hand equalities in Eqs.\ (\ref{eq:daaf}--\ref{eq:dbbf}) are obtained by replacing the first derivative by its value from Eqs.\ (\ref{eq:daf}--\ref{eq:dbf}). In the mixed derivative case there are two possible replacements that give the Maxwell relation $\partial_\alpha\expc{S}=\partial_\beta\expc{R}$.

Once more we can use the self-averaging of $P$ and $Y$ simplify the cumulants; in this case it allows us to express them in terms of the elements of the covariance matrix. For general functions $k_1$ and $k_2$ self-averaging implies that
\begin{multline}
\cum{k_1(P,Y)k_2(P,Y)}=\partial_Pk_1\partial_Pk_2\cum{P^2}\\\qquad+(\partial_Pk_1\partial_Yk_2+\partial_Yk_1\partial_Pk_2)\cum{PY}+\partial_Yk_1\partial_Yk_2\cum{Y^2}\ ,
\end{multline}
where the function on the right-hand-side are evaluated as usual at $P=\bp,~Y=\bp\by$. It follows that
\begin{align}
\cum{R^2}&=(\bp\by)^2\cum{Y^2}\ ,\label{eq:crr}\\
\cum{RS}&=\bp\by Tu'(\bp)^2\cum{P^2}\ ,\\
\cum{S^2}&=T^2u'(\bp)^2\cum{P^2}\ .
\end{align}

Thus, Eqs.\ (\ref{eq:daaf}--\ref{eq:dbbf}) offer an alternative method to calculate three elements of the covariance matrix derived in Sec. \ref{sec:cov}, by taking the parametric derivatives of $R$ and $S$, $\partial \expc{R}=\frac{\bp\by}{T}(\bp\partial\by+\by\partial\bp)$, $\partial S=Tu'(\bp)\partial\bp$, where the `$\partial$' stands for differentiation with respect to either parameter. The parametric derivatives of $\bp$ and $\by$ are obtainable by short calculations starting from Eqs.\ (\ref{eq:bary}) and (\ref{eq:gp}), and give the susceptibilities
\begin{align}
\partial_\beta\bp&=-\frac{\bp\gamma\by}{\eta+\frac{\gamma\by}{1-2\by}}\\
\partial_\beta\by&=\frac{2}{2\by-1}\frac{\partial_\beta\bp}{\bp}\\
\partial_\alpha\bp&=-\frac{\bp^2\by}{\gamma T(2\by-1)}\partial_\beta\bp\\
\partial_\alpha\by&=\frac{1}{\gamma^2(2\by-1)}\Bigl(\frac{2\gamma}{P}\partial_\alpha\bp+\frac{\bp^2}{T}\Bigr)\label{eq:day}
\end{align}
The relevant elements of the covariance matrix are reproduced when Eqs.\ (\ref{eq:crr}--\ref{eq:day}) are substituted in Eqs.\ (\ref{eq:daaf}--\ref{eq:dbbf}).

%\subsection{Lifetime of a superheated pulse}
\subsection{Fluctuations in the cw phase}
For completeness we also present the power fluctuations in the cw phase, where power is evenly divided (on average) between all the laser degrees of freedom. This state is stationary for $\gamma<\gamma^*$ and metastable for all $\gamma>\gamma^*$, although for very large $\gamma$ it is short-lived. The latter case is the self-starting regime \cite{MenyukSelfStarting, Krausz,escape}, and our results hold away from this regime. 

Because the saturable absorption is nonlinear, $F_\text{cw}=f(\bp_\text{cw},0)$, the cw free energy, is independent of $\gamma$ (see Eq.\ (eq:fpy)), and therefore the cw thermodynamics is independent of $\gamma$ \cite{GGF04}. This is also true for the fluctuations, that are therefore (up to small corrections) those of free randomly forced modes subject to the gain saturation constraining potential $u$. The calculation of the covariance matrix is very similar to that of the more complicated mode locked case, and we simply state the results.

The power variance is
\begin{equation}\label{eq:pcw}
\cum{P}_{\text{cw}}=\frac{1}{N|\partial_P^2\phi_\text{cw}(\bp_\text{cw})|}\ .
\end{equation}
Since moments including $Y$ are by definition equal to zero in the cw state Eqs.\ (\ref{eq:cXX}--\ref{eq:cxx'}) together with Eq.\ (\ref{eq:pcw}) determine the remaining nontrivial cumulants $\cum{X_a^2}_\text{cw}$ and $\cum{X_a X'_b}_\text{cw}$.

\section{Conclusions}
The main result of this paper is a complete quantitative characterization of the Gaussian fluctuations in the overall, pulse, and cw power of passively mode locked lasers. It is the first application of SLD to the study of steady-state fluctuations, and further demonstrates the power of this method in the understanding of the interaction of noise and nonlinearity, and its consequences. 
The analysis reveals deeper consequences of the close analogy between a mode locked laser and the equilibrium statistical mechanics of interacting mode systems. In particular the analogy led us to identify and prove fluctuation-dissipation relations for the far-from-equilibrium laser system.

The results were obtained in the framework of a simplified mode locking model. The coarse-grained model provides a good approximation for the behavior of more realistic models and experimental systems, but the quantitative details are model-dependent. Nevertheless, several important conclusions can be drawn from the qualitative properties of our results, that are likely to hold for a wider class of systems. First, we observe that the fluctuations are determined by a further dimensionless, the gain elasticity, in addition to the parameters that determine the steady state (here a single parameter). The gain elasticity measures the response of the amplifier gain to changes in the overall power, and therefore controls the strength of fluctuations in the overall power. As gain elasticity decreases, overall power fluctuations increase, as well as fluctuations in the pulse power and cw power. Secondly, we find that the overall power and the pulse power are always positively correlated, while cw power and pulse power are negatively correlated. This information can in principle be used to predict and correct for pulse fluctuations \emph{before} the arrival of the pulse.  On the other hand, cw power accumulated in different parts of the cavity display positive correlations.

As in all noise driven systems, the fluctuations diverge as the parameters approach the boundary of the region of stability. Close to this boundary, fluctuations cease to be Gaussian; this case can also be studied within the SLD framework \cite{escape}, but with more sophisticated methods than those used here. In particular, the pulse is susceptible to elimination by noise-activation transition to the disordered state. Another potential region of invalidity of the Gaussian distribution, is the neighborhood of a critical point \cite{landau}, absent in the model studied here, but present in more general laser systems \cite{critical-PRL05}. In either case a full analysis of the light fluctuations has to take into account the quantum properties of the spontaneous emission noise \cite{qn}.

The most serious limitation of the static method used here to calculate fluctuation properties is that it can only capture single-time statistics. Many-time statistics, like timing and phase jitter can also be naturally studied within SLD, but, like in statistical mechanics, they require consideration the kinetics of the electromagnetic field, in addition to its steady-state properties. This task is postponed to future study.

\appendix
\section{Steepest-decent evaluation of covariance matrix elements}
Here we outline the evaluation of cumulants $\cum{P^nY^mx^k}$ defined in Sec. \ref{sec:cov}. The expectation values are calculated starting with the definition Eq.\ (\ref{eq:<A>}), fixing the overall power $P$ and first integrating over the $N-2$ degrees other than the cw degree of freedom $\psi_n$, letting $Px=|\psi_n|^2$, and the pulse degree of freedom $\psi_p$, with $NPy=|\psi_p|^2$, obtaining 
\begin{multline}
I_{nmk}=\int dP P^{n+m+k}e^{-Nu(P)}\int dy y^m  e^{N\frac{\alpha}{T}(Py)^2}\\\times\int dx x^k (P(1-y-x/N))^N\ ,
\end{multline}
where the term $(P(1-y-x/N))^N$, the result of the $N-2$ integrations, is equal to the cw-state partition function up to an irrelevant $O(1)$ prefactor \cite{GGF04}. The $x$ integration then leads to Eq.\ (\ref{eq:inmk}) defining  $I_{nmk}$. The remaining integrals over $y$ and $P$ are then evaluated for large $N$ using the standard and well-known steepest-descent method, otherwise known as the Laplace integral method \cite{olver}. It is based on approximating the integrand with its Taylor series near the maximum of the exponential at $(\bar P,\bar y)$. We choose to perform the integration in two steps. The result of the $y$ integration is $I_{nmk}=\int dP \hat I_{nmk}(P)$,
\begin{widetext}
\begin{multline}\label{eq:inmkp}
\hat I_{nmk}(P)=\sqrt{\frac{2\pi}{N|\partial y^2f(P,\by)|}}e^{Nf(P,\by)}P^{n+m+k}\left(\by^m(1-\by)^k+\frac1N\Big(\frac{\partial_y^2(y^m(1-y)^k)|_{\by}}{2|\partial y^2f(P,\by)|}+\cdots\Big)+O\Big(\frac{1}{N^2}\Big)\right)\ .
\end{multline}
where the $\cdots$ stand for several other $O(\frac1N)$ terms that cancel in the evaluation of the cumulants.

The $O(\frac1N)$ term in Eq.\ (\ref{eq:inmkp}) is carried into the next integration, over the $P$ variable, after which $P$ is replaced by $\bp$, and a similar $O(\frac1N)$ is generated giving
\begin{equation}
\tilde I_{nmk}=\frac{\partial_y^2(y^m(1-y)^k)|_{\by(\bp)}}{2|\partial y^2f(\bp,\by)|}+\frac{\partial_P(P^{n+m+k}\by(P)^m(1-\by(P))^k)|_{\bp}}{2|\partial_P^2\phi(\bp)|}\ .
\end{equation}
\end{widetext}

\end{document}